# Space charge impedances of a rectangular beam with longitudinal density modulations inside a rectangular chamber

Yingjie Li[a,*], Lanfa Wang[b]

[a]Department of Physics, Michigan State University, East Lansing, MI 48824, USA
[b]SLAC National Accelerator Laboratory, Menlo Park, CA 94025, USA

**Abstract**

This paper studies the space charge impedances of a rectangular beam inside a rectangular chamber, and the limiting case, e.g., a rectangular beam between parallel plates, respectively. The charged beam has uniform density in vertical direction and arbitrary distribution in horizontal direction. The method of separation of variables is used to calculate the space charge potentials, fields, and impedances which are valid in the whole perturbation wavelength spectrum. Comparisons between the theoretical calculations and the numerical simulations are also provided and they match quite well. It is shown that the rectangular beam shape may help to reduce the longitudinal space charge impedances.

## 1. Introduction

In order to study the longitudinal beam instabilities due to the interactions between the perturbed beam and the conducting vacuum chamber surrounding it, various space charge field models with different cross-sections of the beam and chamber have been investigated. Ref. [1] derived the longitudinal space charge impedances of a round beam inside a rectangular chamber in the long-wavelength limits. Ref. [2] and Ref. [3] studied the longitudinal resistive-wall instability and the space-charge driven microwave instability, respectively, using a model consisting of a round beam inside a round vacuum chamber. In the research for future linear colliders, some merits of flat (planar or rectangular) electromagnetic structures have been found and aroused the interests of beam physicists, such as the reduced space charge forces [4] [5]. Ref. [6] explored the properties of a planar beam between a pair of perfectly conducting plates. It concluded that, comparing with the conventional axially symmetric configurations, the flat geometries of both the beam and the chamber may help to reduce the longitudinal space charge fields. The two-dimensional (2D) electrostatic space charge field of a rectangular beam inside a rectangular chamber was solved by Ref. [7] using the method of separation of variables. While in this model, the field was induced only by the unperturbed (constant) beam intensity without longitudinal modulations. The results are only valid when the perturbation wavelengths of the longitudinal charge density are much larger than the transverse dimensions of vacuum chamber, hence cannot be used directly in the study of short-wavelength instabilities, such as microwave instability and micro-bunching instability. Another model of rectangular beam inside rectangular chamber in Ref. [8] assumed the beam perturbation took place in the vertical direction, and hence this model was devoted to study the transverse resistive wall instability instead of the longitudinal one.

This paper introduced a three-dimensional (3D) space charge field model consisting of a rectangular beam with sinusoidal longitudinal density modulations inside a rectangular vacuum chamber. The vertical charge density is assumed to be uniform, while the horizontal beam distributions are not restricted. By applying the Fourier expansion to the horizontal distributions and using the method of separation of variables, the space charge potentials and fields within the chamber can be solved analytically in Cartesian coordinate system. The results are valid in the whole perturbation wavelength spectrum and can be used to study the microwave instability. The longitudinal space charge impedances of this model and its limiting case of parallel plate model were derived for the convenience of beam instability analysis. A general-purpose simulation code based on the Finite Element Method (FEM) was developed by us. The theoretical longitudinal space charge impedances are consistent with the numerical simulation results quite well. The effects of the different beam and chamber dimensions on the space charge impedances were investigated.

* Corresponding author.
*Email address*: liyingji@msu.edu (Yingjie Li).

This paper was organized as follows. Section 2 briefly introduced the space charge field model. Section 3 derived the analytical solutions to the space charge potentials and fields of this model. Section 4 derived the longitudinal space charge impedances of a rectangular beam inside a rectangular chamber, and a rectangular beam between parallel plates, respectively. Section 5 provided the case studies of the longitudinal space charge impedances using both simulation and theoretical methods. The effects of beam and chamber dimensions on impedances were explored by the method of scanning.

## 2. Field model of a rectangular beam inside a rectangular chamber

The geometry of the cross-section of a rectangular beam inside a grounded, perfectly conducting vacuum chamber is shown in Fig. 1. The beam and the chamber are coaxial with the center located at ($w$, 0). The full width and height of the inner boundary of the chamber are $2w$ and $2h$, respectively. The full width and height of the beam are $2a$ and $2b$, respectively. The horizontal beam dimension $2a$ is variable and can be as wide as the full chamber width $2w$.

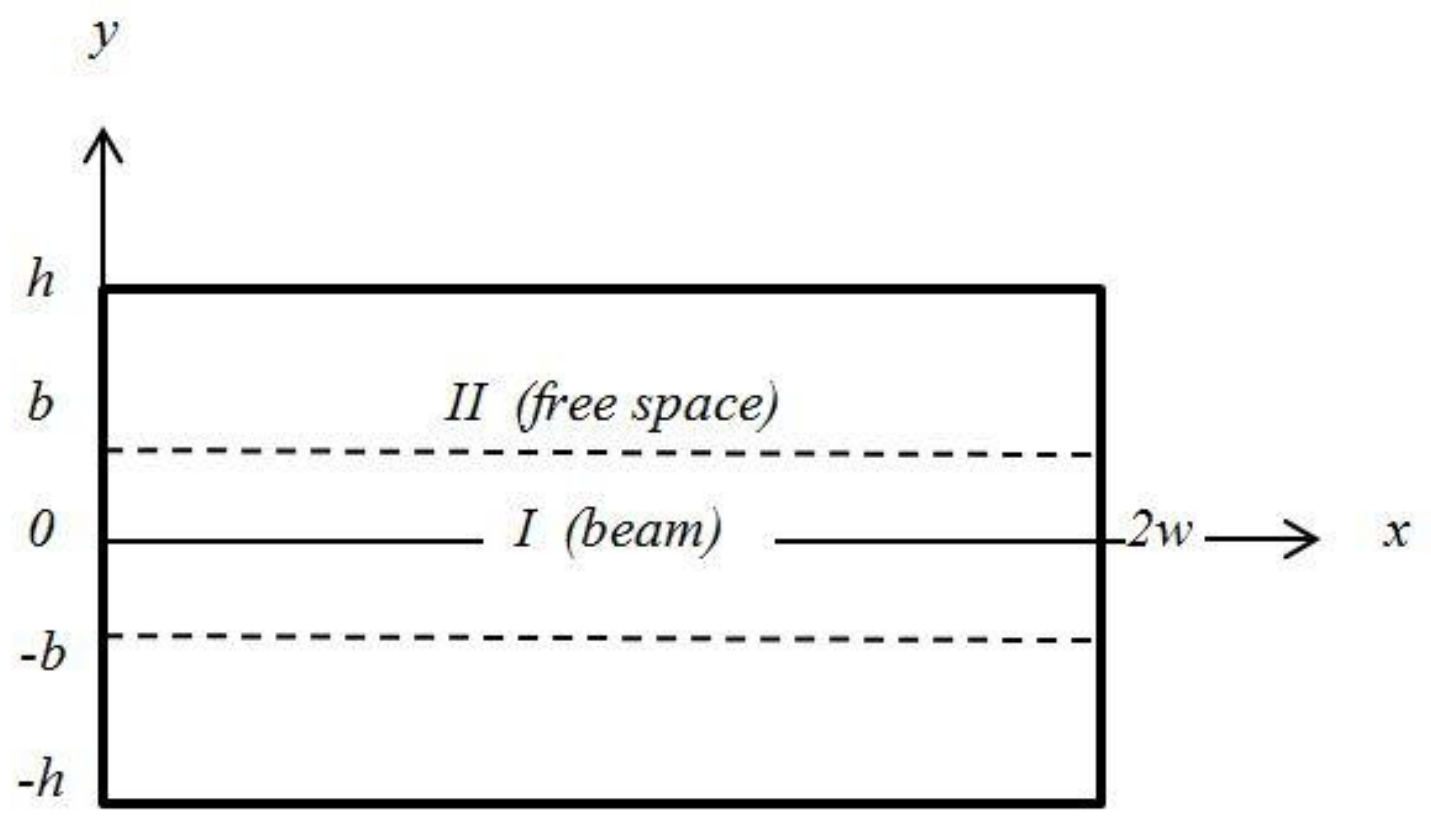

Fig. 1. A rectangular beam inside a rectangular chamber.

Assume the vertical particle distribution is uniform in the region of $-b \le y \le -b$. For the longitudinal charge distributions, since the unperturbed charge density $\Lambda_0$ does not affect the longitudinal space charge fields, we can only keep the perturbed charge density components.

In the *lab* frame, assume the line charge density and beam current have sinusoidal modulations along the longitudinal coordinate $z$, and can be written in the form of propagating waves as

$$\Lambda(z,t) = \Lambda_k \exp[i(kz - \omega t)], \qquad I(z,t) = I_k \exp[i(kz - \omega t)]. \tag{1}$$

respectively, where $\Lambda_k$ and $I_k$ are the amplitudes, $I_k = \Lambda_k \beta c$, β is the relativistic speed of the beam, $c$ is the speed of light, $\omega$ is the angular frequency of the perturbations, $k$ is the wave number of the line charge density modulations.

In order to calculate the longitudinal space charge fiend inside the beam in the *lab* frame, first, we can calculate the electrostatic potentials and fields in the *rest* frame of the beam, and then convert them into the lab frame by Lorentz transformation.

In the *rest* frame, the line charge density of a beam can be simplified as

$$\Lambda'(z') = \Lambda'_k \cos(k' z'), \tag{2}$$

where the symbol prime stands for the *rest* frame.

For general purpose, we assume there are no restrictions for the horizontal beam distributions within the chamber. If the dependence of the perturbed volume charge density $\rho'(x', y', z')$ on $x'$ in the *rest* frame can be described by a function of $G(x')$, then

$$\rho'(x',y',z') = \begin{cases} \dfrac{\Lambda'_k \cos(k'z')}{2b} G(x'), & |y'| \le b. \\ 0, & b < |y'| \le h. \end{cases} \quad (3)$$

where $G(x')$ satisfies the normalization condition of

$$\int_0^{2w} G(x')dx' = 1, \quad (4)$$

and the volume charge density correlates with the line charge density

$$\int_{-h}^{h} dy' \int_0^{2w} \rho'(x',y',z')dx' = \Lambda'(z'). \quad (5)$$

In order to solve the Poisson equation in the Cartesian coordinate system analytically and conveniently using the method of separation of variables, the normalized horizontal distribution function $G(x')$ can be written as a Fourier series. Since the charge must vanish on the chamber side walls at $x' = 0$ and $x' = 2w$, we can expand $G(x')$ to a sinusoidal series

$$G(x') = \frac{1}{2w}\sum_{n=1}^{\infty} g_n' \sin(\eta_n x'), \quad (6)$$

$$\eta_n = \frac{n\pi}{2w}. \quad (7)$$

The dimensionless Fourier coefficient $g_n'$ can be calculated by

$$g_n' = 2\int_0^{2w} G(x')\sin(\eta_n x')dx'. \quad (8)$$

From Eq. (3) and Eq. (6), the volume charge density in the *rest* frame can be expressed as

$$\rho'(x',y',z') = \begin{cases} \dfrac{\Lambda_k' \cos(k'z')}{4bw} \displaystyle\sum_{n=1}^{\infty} g_n' \sin(\eta_n x'), & |y'| \le b. \\ 0, & b < |y'| \le h. \end{cases} \quad (9)$$

## 3. Calculation of the space charge potentials and fields

In Region *I* (charge region) and Region *II* (free space region), the electrostatic space charge potentials $\varphi_I'(x', y', z')$ and $\varphi_{II}'(x', y', z')$ in the *rest* frame satisfy the Poisson equation and Laplace equation, respectively. Then we have

$$\left(\frac{\partial^2}{\partial x'^2} + \frac{\partial^2}{\partial y'^2} + \frac{\partial^2}{\partial z'^2}\right)\varphi_I'(x',y',z') = -\frac{\Lambda_k' \cos(k'z')}{4\varepsilon_0 bw}\sum_{n=1}^{\infty} g_n' \sin(\eta_n x'), \quad (10)$$

$$\left(\frac{\partial^2}{\partial x'^2} + \frac{\partial^2}{\partial y'^2} + \frac{\partial^2}{\partial z'^2}\right)\varphi_{II}'(x',y',z') = 0, \quad (11)$$

where $\varepsilon_0 = 8.85\times10^{-12}$ F/m is the permittivity in free space.

The basic components of the solutions to Eq. (11) and the homogeneous form of Eq. (10) can be written as

$$\varphi_h' = X(x')Y(y')\cos(k'z'). \tag{12}$$

The possible configurations of the solutions to *X(x')* and *Y(y')* may have the forms of

$$X(x') \sim \cos(\eta_n x'), \sin(\eta_n y') \text{ or their combinations,} \tag{13}$$

and

$$Y(y') \sim \cosh(v_n' y'), \sinh(v_n' y') \text{ or their combinations,} \tag{14}$$

respectively, where $$v_n'^2 = \eta_n^2 + k'^2, \quad n=1, 2, 3 \ldots\ldots \tag{15}$$

Considering the boundary conditions (a) $\varphi' = 0$, $E_y' = 0$ at $x' = 0, 2w$; (b) $\varphi' = 0$, $E_x' = 0$ at $y' = \pm h$, and the potential $\varphi'(x', y', z')$ should be even functions of $y'$, the basic components of solutions to Eq. (11) and the homogeneous form of Eq. (10) may have the following forms

In region *I* (charge region): $$\varphi_{h,I}' \sim \sin(\eta_n x')\cosh(v_n' y')\cos(k'z'), \tag{16}$$

In region *II* (free space region): $$\varphi_{h,II}' \sim \sin(\eta_n x')\sinh[v_n'(h - |y'|)]\cos(k'z'). \tag{17}$$

The particular solution to the inhomogeneous Eq. (10) can be written as

$$\varphi_{i,I}'(x', y', z') = \cos(k'z')\sum_{n=1}^{\infty} C_n' \sin(\eta_n x'). \tag{18}$$

Plugging Eq. (18) into Eq. (10) and comparing the coefficients of the like terms of the two sides gives the coefficients $C_n'$

$$C_n' = \frac{\Lambda_k' g_n'}{4\varepsilon_0 b w v_n'^2}. \tag{19}$$

Then in region *I* (charge region), the field potentials in the *rest* frame are

$$\varphi_I'(x', y', z') = \varphi_{h,I}' + \varphi_{i,I}' = \cos(k'z')\sum_{n=1}^{\infty} \sin(\eta_n x')[A_n'\cosh(v_n' y') + C_n']. \tag{20}$$

In region *II* (free space region), the field potentials in the *rest* frame are

$$\varphi_{II}'(x', y', z') = \cos(k'z')\sum_{n=1}^{\infty} B_n' \sin(\eta_n x')\sinh[v_n'(h - |y'|)]. \tag{21}$$

The boundary conditions between Region *I* and Region *II* are: at $y' = \pm b$, $\varphi_I' = \varphi_{II}'$, $\partial\varphi_I'/\partial y' = \partial\varphi_{II}'/\partial y'$. Then the coefficients $A_n'$ and $B_n'$ can be determined as

$$A_n' = -\frac{\cosh[v_n'(h-b)]}{\cosh(v_n' h)} C_n', \quad B_n' = \frac{\sinh(v_n' b)}{\cosh(v_n' h)} C_n'. \tag{22}$$

Finally, the space charge potentials in the *rest* frame are

(a) In region *I* (charge region), $0 \le |y'| \le b$,

$$\varphi_I'(x', y', z') = \frac{\Lambda_k' \cos(k'z')}{4\varepsilon_0 b w}\sum_{n=1}^{\infty} \frac{g_n'}{v_n'^2}\sin(\eta_n x')\{1 - \frac{\cosh[v_n'(h-b)]}{\cosh(v_n' h)}\cosh(v_n' y')\}. \tag{23}$$

(b) In region *II* (free space), $b < |y'| \le h$,

$$\varphi_{II}'(x',y',z') = \frac{\Lambda_k' \cos(k'z')}{4\varepsilon_0 bw} \sum_{n=1}^{\infty} \frac{g_n'}{\nu_n'^2} \frac{\sinh(\nu_n' b)}{\cosh(\nu_n' h)} \sin(\eta_n x') \sinh[\nu_n'(h - |y'|)]. \tag{24}$$

For a beam with rectangular cross-section and uniform transverse charge density, the volume charge density in the *rest* frame can be expressed as

$$\rho'(x',y',z') = \begin{cases} \dfrac{\Lambda_k'}{4ab} \cos(k'z'), & w-a \le x' \le w+a, |y'| \le b. \\ 0, & x' < w-a, x' > w+a, b < |y'| \le h. \end{cases} \tag{25}$$

Comparing Eq. (25) with Eq. (3) gives *G(x')* is equal to *1/2a* inside the beam and 0 outside of the beam, respectively. Then $g_n'$ can be calculated from Eq. (8) as

$$g_n' = \frac{2}{\eta_n a} \sin(\eta_n w) \sin(\eta_n a). \tag{26}$$

inside the beam and 0 outside of the beam, respectively.

According to Eq. (23), the longitudinal space charge field inside the beam in the *rest* frame can be calculated as

$$E_{z,I}'(x',y',z') = -\frac{\partial \varphi_I'(x',y',z')}{\partial z'} = -\frac{\dfrac{d\Lambda'(z')}{dz'}}{4\varepsilon_0 bw} \sum_{n=1}^{\infty} \frac{g_n'}{\nu_n'^2} \sin(\eta_n x') \{1 - \frac{\cosh[\nu_n'(h-b)]}{\cosh(\nu_n' h)} \cosh(\nu_n' y')\}. \tag{27}$$

According to the theory of relativity, the relations of parameters between the *rest* frame and the *lab* frame are

(a) The longitudinal electric field is invariant, i.e.,

$$E_{z,I}' = E_{z,I}\,, \tag{28}$$

(b) The wave number $\quad k' = k/\gamma, \quad$ (29)

(c) The coordinates $\quad x' = x, \quad y' = y, \quad z' = \gamma(z - \beta ct), \quad$ (30)

(d) The line charge density amplitude $\quad \Lambda_k' = \Lambda_k / \gamma\,, \quad$ (31)

(e)

$$\nu_n'^2 = \eta_n^2 + k'^2 = \eta_n^2 + \frac{k^2}{\gamma^2}, \tag{32}$$

(f)

$$\frac{d\Lambda'(z)}{dz'} = -k'\Lambda_k' \sin(k'z') = -\frac{k\Lambda_k}{\gamma^2} \sin(kz - \omega t) \tag{33}$$

If we choose exponential representation as used in Eq. (1), Eq. (33) can also be expressed as

$$\frac{d\Lambda'(z)}{dz'} = \frac{1}{\gamma^2} \frac{\partial \Lambda(z,t)}{\partial z} \tag{34}$$

where $\gamma$ is the relativistic factor, the subscript *lab* stands for the lab frame. Then the longitudinal electric field in the *lab* frame becomes

$$E_{z,I}(x,y,z,t) = -\frac{\dfrac{\partial \Lambda(z,t)}{\partial z}}{4\varepsilon_0 bw\gamma^2} \sum_{n=1}^{\infty} \frac{g_n}{\nu_n^2} \sin(\eta_n x) \{1 - \frac{\cosh[\nu_n(h-b)]}{\cosh(\nu_n h)} \cosh(\nu_n y)\}. \tag{35}$$

where

$$g_n = g_n' = \frac{2}{\eta_n a}\sin(\eta_n w)\sin(\eta_n a), \tag{36}$$

$$\nu_n^{\ 2} = \nu'^2_n = \eta_n^{\ 2} + k'^2 = \eta_n^{\ 2} + \frac{k^2}{\gamma^2}. \tag{37}$$

## 4. Longitudinal space charge impedances

The average longitudinal electric fields over the cross-section of the beam at *z* and time t are

$$< E_{z,I}(z,t) >= \frac{1}{4ab}\int_{-b}^{b} dy \int_{w-a}^{w+a} E_{z,I}(x,y,z,t)dx$$

$$= -\frac{\frac{\partial\Lambda(z,t)}{\partial z}}{4\varepsilon_0 bw\gamma^2}\sum_{n=1}^{\infty}\frac{g_n}{\nu_n^{\ 2}} < \sin(\eta_n x) > \{1 - \frac{\cosh[\nu_n(h-b)]}{\cosh(\nu_n h)} < \cosh(\nu_n y) >\}, \tag{38}$$

where

$$< \sin(\eta_n x) >= \frac{1}{2a}\int_{w-a}^{w+a}\sin(\eta_n x)dx = \frac{1}{2}g_n, \tag{39}$$

$$< \cos(\nu_n y) >= \frac{1}{2b}\int_{-b}^{b}\cosh(\nu_n y)dy = \frac{1}{b\nu_n}\sinh(\nu_n b). \tag{40}$$

Finally, the average longitudinal space charge fields in the beam region can be expressed as

$$< E_{z,I}(z,t) >= -\frac{1}{4\varepsilon_0 bw\gamma^2}\chi(k)\frac{\partial\Lambda(z,t)}{\partial z}, \tag{41}$$

where

$$\chi(k) = \sum_{n=1}^{\infty}\frac{g_n^{\ 2}}{2\nu_n^{\ 2}}\{1 - \frac{\cosh[\nu_n(h-b)]}{\nu_n b\cosh(\nu_n h)}\sinh(\nu_n b)\}. \tag{42}$$

The sum of the infinite series in Eq. (42) can be evaluated by truncating it to a finite number of terms, as long as the sum converges well.

The energy loss per turn of a unit charge in a storage ring due to the longitudinal space charge field is

$$- < E_{z,I}(z,t) > C_0 = Z^{\parallel}_{0,sc}(k)I_k \exp[i(kz - \omega t)], \tag{43}$$

where $C_0$ is the circumference of the storage ring, $Z^{\parallel}_{0,sc}(k)$ is the longitudinal space charge impedance of the rectangular beam inside the rectangular chamber. It is easy to get from Eqs. (1)(41) and (43) that the impedance per unit length ($\Omega$ /m) is

$$\frac{Z^{\parallel}_{0,sc}(k)}{C_0} = i\frac{Z_0 k}{4\beta bw\gamma^2}\chi(k), \tag{44}$$

where $Z_0$ = 377 $\Omega$ is the impedance of free space, $R$ is the average radius of the storage ring. If the impedance is evaluated by the longitudinal space charge fields on the beam axis ($w$, 0), since in Eq. (35), $sin(\eta_n x) = sin(n\pi/2)$, $cosh(\nu_n y)$ = 1, then $\chi(k)$ in Eq. (42) should be replaced by

$$\chi_{axis}(k) = \sum_{n=1}^{\infty} \frac{g_n}{{\nu_n}^2} \sin(\frac{n\pi}{2})\{1 - \frac{\cosh[\nu_n(h-b)]}{\cosh(\nu_n h)}\}. \tag{45}$$

For a special case of infinite $h$, i.e., the rectangular chamber becomes a pair of vertical parallel plates separated by $2w$, since $\lim_{h\to\infty}\{\cosh[v_n(h-b)]/\cosh(v_n h)\} = \cosh(v_n b) - \sin(v_n b)$, the parameter $\chi(k)$ in Eq. (42) can be simplified as

$$\chi_{vpp}(k) = \sum_{n=1}^{\infty} \frac{{g_n}^2}{2{\nu_n}^2}[1 - \frac{\cosh(\nu_n b) - \sinh(\nu_n b)}{\nu_n b} \sinh(\nu_n b)]. \tag{46}$$

Eqs. (44) (46) give the longitudinal space charge impedances per unit length of a rectangular beam between a pair of vertical parallel plates separated by $2w$. In Eq. (46), if $b$ is infinite, i.e. a rectangular beam with infinite height between two vertically parallel plates, since the last part in the right hand side of Eq. (46) becomes zero, then

$$\chi_{vpp,b=\infty}(k) = \sum_{n=1}^{\infty} \frac{{g_n}^2}{2{\nu_n}^2}. \tag{47}$$

For a special case of $w\to\infty$, i.e., the rectangular chamber becomes a pair of horizontal parallel plates separated by $2h$, if we make exchanges $a\leftrightarrow b$, $w\leftrightarrow h$, it is easy to get its impedances from Eqs. (44)(46) that

$$\frac{Z^{\parallel}_{0,hpp}(k)}{C_0} = i\frac{Z_0 k}{4\beta a h \gamma^2}\chi_{hpp}(k), \tag{48}$$

$$\chi_{hpp}(k) = \sum_{n=1}^{\infty} \frac{{g_{n,hpp}}^2}{2{\nu_{n,hpp}}^2}[1 - \frac{\cosh(\nu_{n,hpp} a) - \sinh(\nu_{n,hpp} a)}{\nu_{n,hpp} a} \sinh(\nu_{n,hpp} a)], \tag{49}$$

where

$$\eta_{n,hpp} = \frac{n\pi}{2h}, \tag{50}$$

$${\nu_{n,hpp}}^2 = {\eta_{n,hpp}}^2 + \frac{k^2}{\gamma^2}, \quad n=1,\ 2,\ 3\ \ldots\ldots \tag{51}$$

$$g_{n,hpp} = \frac{2}{\eta_{n,hpp} b}\sin(\eta_{n,hpp} h)\sin(\eta_{n,hpp} b). \tag{52}$$

Eqs. (48-52) give the longitudinal space charge impedances of a rectangular beam between a pair of horizontal parallel plates separated by $2h$. In Eq. (49), if $a\to\infty$, i.e. a rectangular beam with infinite width between two horizontal parallel plates, since $[cosh(\nu_{n,hpp}a)-sinh(\nu_{n,hpp}a)]sinh(\nu_{n,hpp}\,a)/\nu_{n,hpp}a \to 0$, then

$$\chi_{hpp,a=\infty}(k) = \sum_{n=1}^{\infty} \frac{{g_{n,hpp}}^2}{2{\nu_{n,hpp}}^2}. \tag{53}$$

## 5. Case studies of the longitudinal space charge impedances

We developed a simulation code that can solve the Poisson equation numerically based on the Finite Element Method (FEM) [9]. The code can be used to calculate the space charge potentials, fields and impedances of the beam-chamber system with any configurations of the charge distributions and boundary shapes. In the *rest* frame, assume the harmonic volume charge density can be written as product of the transverse and longitudinal components

$$\rho'(x',y',z')=\rho_{\perp}'(x',y')\Lambda'(z')=\rho_{\perp}'(x',y')\Lambda_k' e^{ik'z'}. \tag{54}$$

where $\int \rho_{\perp}'(x',y')dx'dy'=1.$ Similarly, the potential due to the harmonic charge density is written as

$$\varphi'(x',y',z')=\varphi_{\perp}'(x',y')e^{ik'z'}. \tag{55}$$

The Poisson equation with Eqs. (54) and (55) becomes

$$(\nabla'^2_{\perp}-k'^2)\varphi_{\perp}'=-\Lambda_k'\frac{\rho_{\perp}'(x',y')}{\varepsilon_0}. \tag{56}$$

where $\nabla'^2_{\perp}=\partial^2/\partial x'^2+\partial^2/\partial y'^2$ and $\varphi_{\perp}'=0$ on the metal boundary. The potentials given by Eq. (56) with arbitrary beam and chamber shapes can be solved using FEM. The whole domain is first divided into many small element regions (finite element). For each element, the strong form of the Poisson equation Eq. (56) can be rewritten as the FEM equation

$$\mathbf{M}\varphi_{\perp}'+k'^2\mathbf{B}=\mathbf{Q}, \tag{57}$$

where

$$M^e_{ij}=\iint_{S^e}\left(\frac{\partial N_i}{\partial x'}\frac{\partial N_j}{\partial x'}+\frac{\partial N_i}{\partial y'}\frac{\partial N_j}{\partial y'}\right)dx'dy', \tag{58}$$

$$B^e_i=\iint_{S^e}N_iN_j dx'dy', \tag{59}$$

$$Q^e_i=\frac{q_i}{\varepsilon_0}. \tag{60}$$

Here *N(x', y')* is called the shape function in FEM, by which the potentials at field point *P(x', y')* within an element can be interpolated by the potentials of its neighboring nodes, it is related to the coordinates of the field point *P(x', y')* and the nodes of the element region. ***M*** is the stiffness matrix with matrix element $M^e_{i,j}$, *i* and *j* are the node indices of the finite element, $S^e$ is the integration boundary of the finite element, $q_i$ is the charge at the node *i*, which is proportional to the harmonic line charge density amplitude $\Lambda_k'$. The $\varphi_{\perp}'$ of Eq. (57) at all nodes satisfying equations Eqs. (57)-(60) and the boundary condition $\varphi_{\perp}' = 0$ on the chamber wall can be solved numerically. Then the total potentials in the *rest* frame can be calculated from Eq. (55), the corresponding longitudinal space charge fields and impedances in the *lab* frame can be calculated using the similar procedures in Sect. 4.

Now we can use the rectangular beam and chamber model to estimate the longitudinal space charge impedances of the coasting $H_2^+$ beam in the Small Isochronous Ring (SIR) at Michigan State University (MSU) [10]. The ring circumference is $C_0 = 6.58$ m, the kinetic energy of the beam is $E_k = 20$ keV ($\beta \approx 0.0046$, $\gamma \approx 1.0$), the cross-section of the vacuum chamber is rectangular with $w = 5.7$ cm, $h = 2.4$ cm, the real beam is approximately round with radius $r_0 = 0.5$ cm. We can use a square beam model with $a = b = r_0 = 0.5$ cm to mimic the round beam.

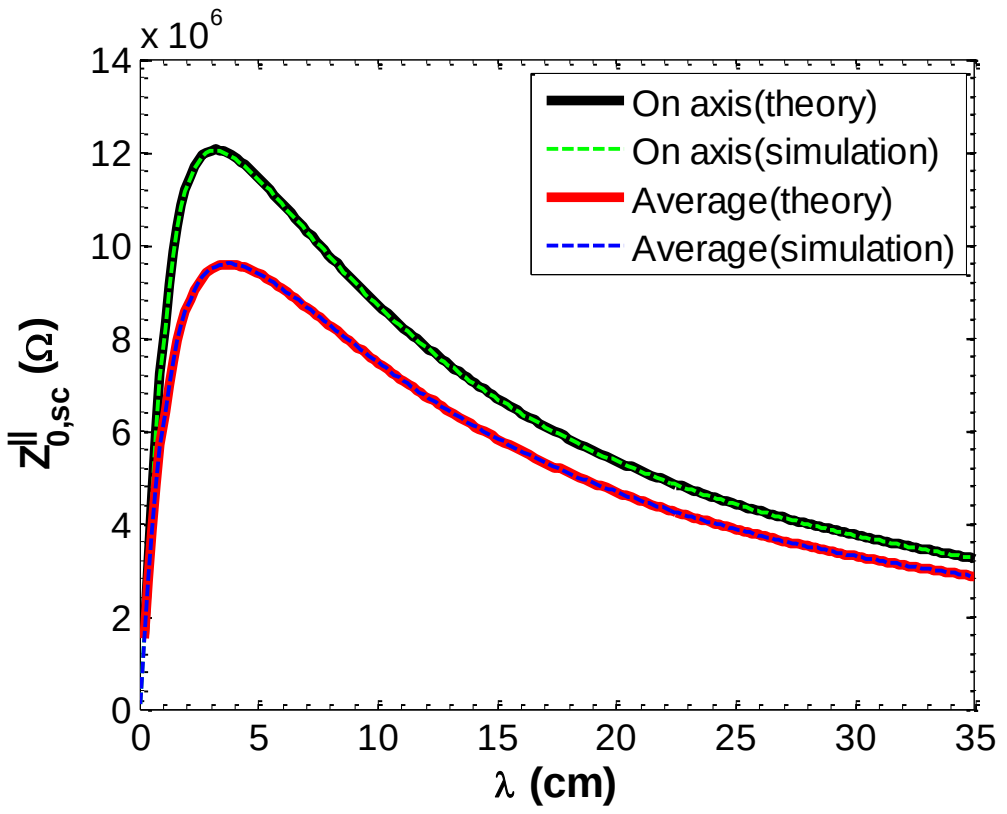

Fig. 2. Comparisons of the on-axis and average longitudinal space charge impedances between the theoretical calculations and numerical simulations for a beam model of square cross-section inside rectangular chamber with $w$ = 5.7 cm, $h$ =2.4 cm, $a = b =$ 0.5 cm.

Fig. 2 shows the comparisons of the on-axis and average longitudinal space charge impedances of SIR beam between the theoretical calculations and numerical simulations using a square beam model. We can see that the theoretical and simulated impedances match quite well. Note that the impedances evaluated by the longitudinal electric fields on the beam axis are higher than those averaged over the beam cross-section, it may overestimate the longitudinal space charge effects. For this reason, we only plot the impedances averaged over the beam cross-section in the following figures.

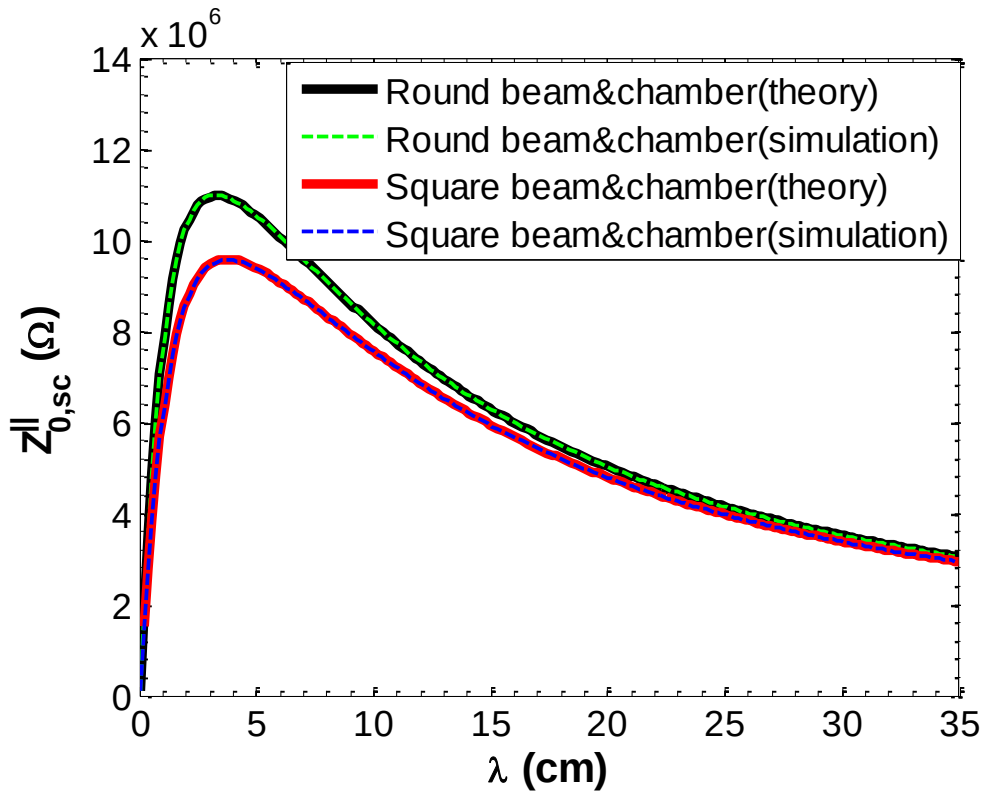

Fig. 3. Comparisons of the longitudinal space charge impedances between the square and round models ($w = h = r_w$ = 3.0 cm, $a = b = r_0$ = 0.5 cm).

Fig. 3 shows the comparisons of the longitudinal space charge impedances between the square and round field models. The longitudinal space charge impedances per unit length of a round beam of radius $r_0$ inside a round chamber of radius $r_w$ can be derived from Ref. [3] as

$$\frac{Z^{\|}_{0,rr}(\bar{k})}{C_0} = i\frac{Z_0\bar{k}}{\beta\pi\gamma}\chi_{rr}(\bar{k}), \tag{61}$$

$$\chi_{rr}(\bar{k}) = \frac{1}{(\bar{k}r_0)^2} - \frac{\langle I_0(\bar{k}r)\rangle}{\bar{k}r_0 I_0(\bar{k}r_w)}[K_1(\bar{k}r_0)I_0(\bar{k}r_w) + K_0(\bar{k}r_w)I_1(\bar{k}r_0)], \tag{62}$$

where $\bar{k} = k/\gamma$, $I_0(x)$, $I_1(x)$, $K_0(x)$, $K_1(x)$ are the modified Bessel functions, and

$$\langle I_0(\bar{k}r)\rangle = \frac{1}{\pi r_0^2}\int_0^{2\pi} d\theta \int_0^{r_0} I_0(\bar{k}r) r dr. \tag{63}$$

The parameters used in the calculations are $w = h = r_w = 3.0$ cm, $a = b = r_0 = 0.5$ cm. We can observe that the model with square beam and chamber shapes has lower longitudinal space charge impedances compared with the round one. At large perturbation wavelengths, the impedances of the two field models are close to each other.

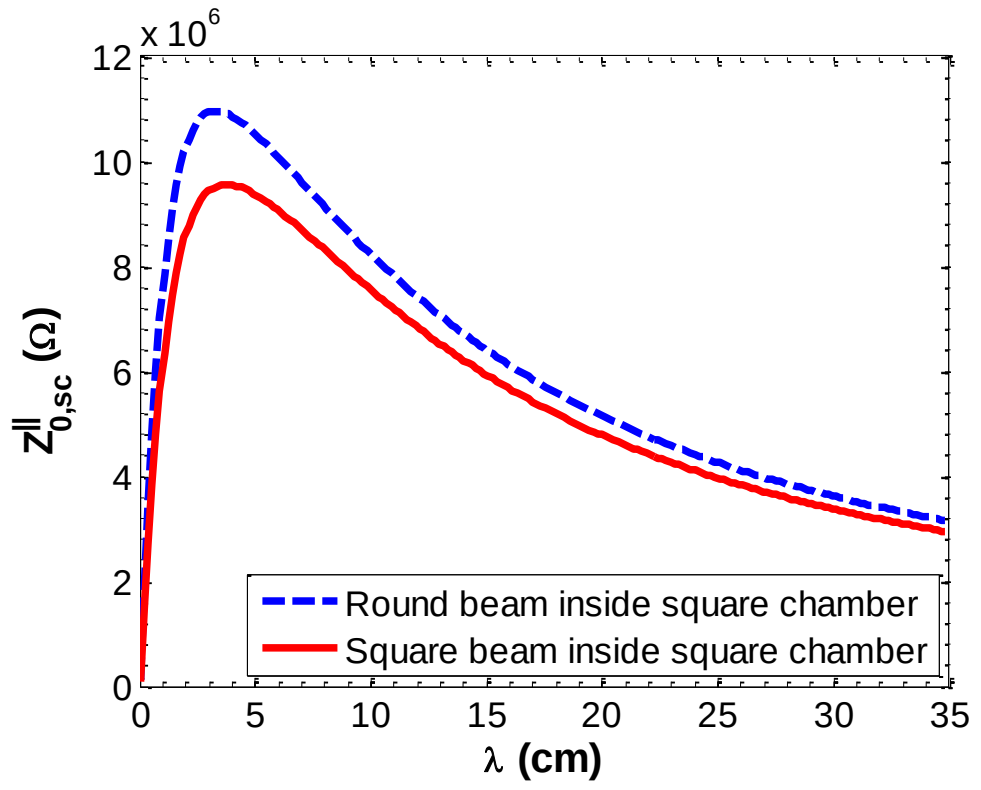

Fig. 4. Simulated longitudinal space charge impedances of square and round beam models in square chamber ($w = h = 3.0$ cm, $a = b = r_0 = 0.5$ cm), respectively.

Fig. 4 shows the simulated longitudinal space charge impedances of the square and round $H_2^+$ beam of 20 keV inside the same square chamber. The parameters used in the calculations are $w = h = 3.0$ cm, $a = b = r_0 = 0.5$ cm. We can observe that the square beam has relatively lower longitudinal space charge impedances than the round beam. The difference of impedances is caused by the different beam shapes. For a beam with fixed line charge density, the square beam has a larger area of cross-section than the round beam inscribing it, hence has smaller volume charge density, lower longitudinal electric fields and impedances. At large perturbation wavelengths, the impedances of the two field models are close to each other.

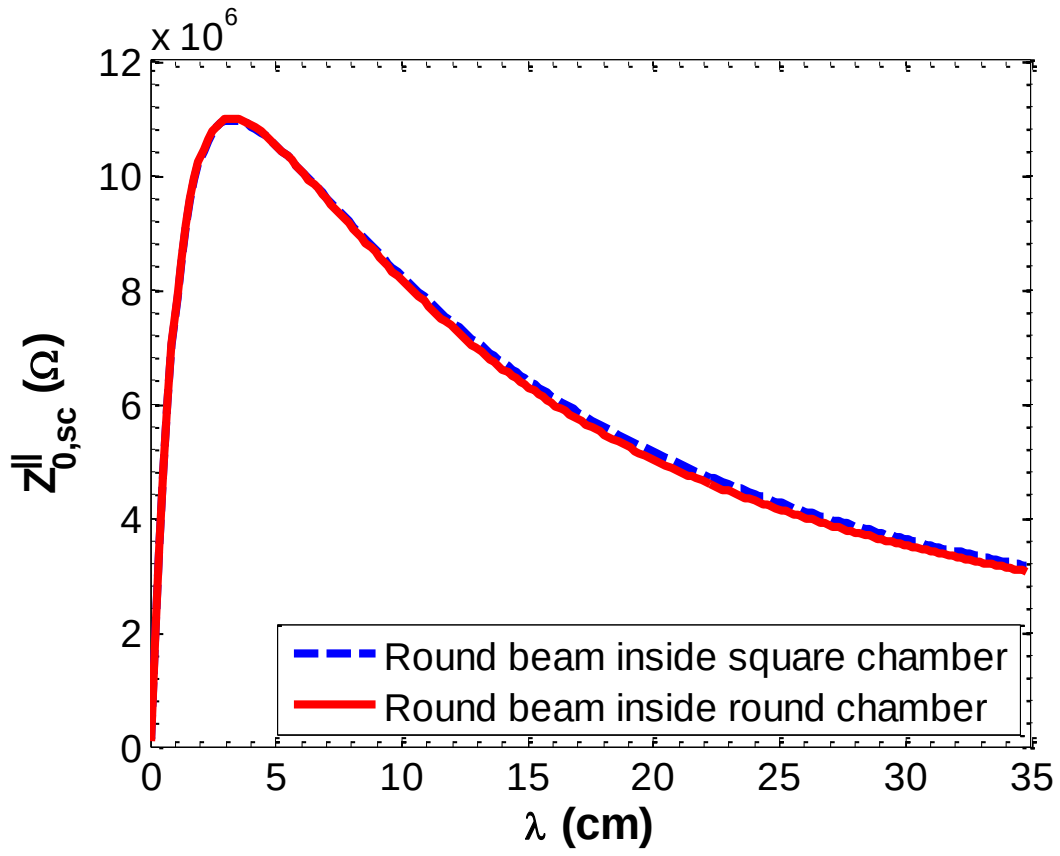

Fig. 5. Simulated longitudinal space charge impedances of round beam inside square and round chambers ($w = h = r_w = 3.0$ cm, $r_0 = 0.5$ cm), respectively.

Fig. 5 shows the simulated longitudinal space charge impedances of a round $H_2^+$ beam of 20 keV inside the round and square chambers, respectively. The parameters used in the calculations are $w = h = r_w = 3.0$ cm, $r_0 = 0.5$ cm. We can observe that the two curves are close to each other, and the square chamber model has relatively higher longitudinal space charge impedances than the round chamber model. The reason for this tiny difference is that the four corners of the square chamber are relatively farther away from the beam axis compared with a round chamber inscribing the square chamber, thus the shielding effects of the square chamber due to image charges are weaker, and therefore the longitudinal space charge fields become stronger. At large perturbation wavelengths, the impedances of the two field models are close to each other.

Figs. 3-5 show that the lower impedances of the rectangular beam and chamber model in Fig. 3 mainly originate from the different beam shapes rather than the chamber shapes.

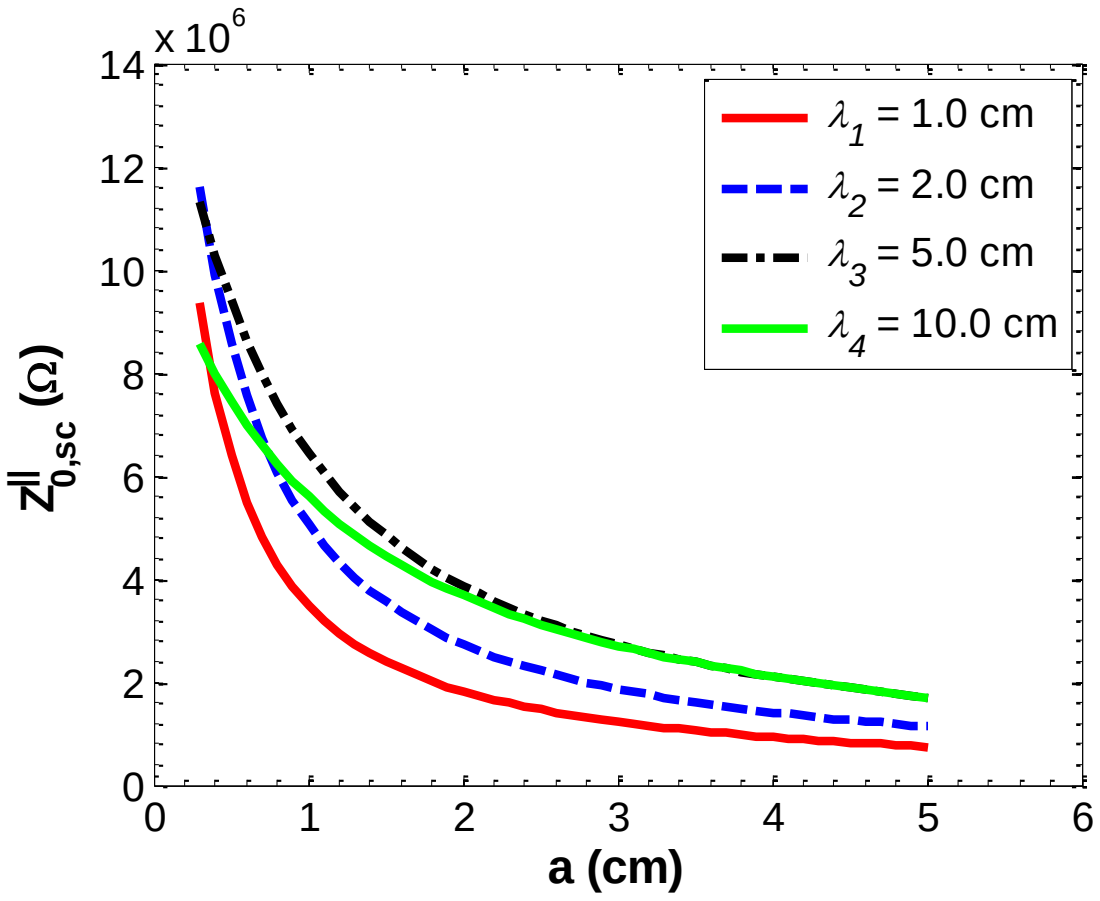

Fig. 6. Longitudinal space charge impedances of rectangular beam model with different half widths *a* inside rectangular chamber (*w* = 5.7 cm, *h* = 2.4 cm, *a* is variable, *b* = 0.5 cm).

Fig. 6 shows the calculated longitudinal space charge impedances of four perturbation wavelengths for a 20 keV $H_2^+$ beam model of rectangular cross-section inside the rectangular chamber of SIR. The parameters used in the calculations are $w$ = 5.7 cm, $h$ = 2.4 cm, $b$ = 0.5 cm, the half beam width $a$ is variable. We can see the longitudinal space charge impedances decrease with beam width *2a* for a fixed beam height *2b* due to dilutions of the volume charge densities.

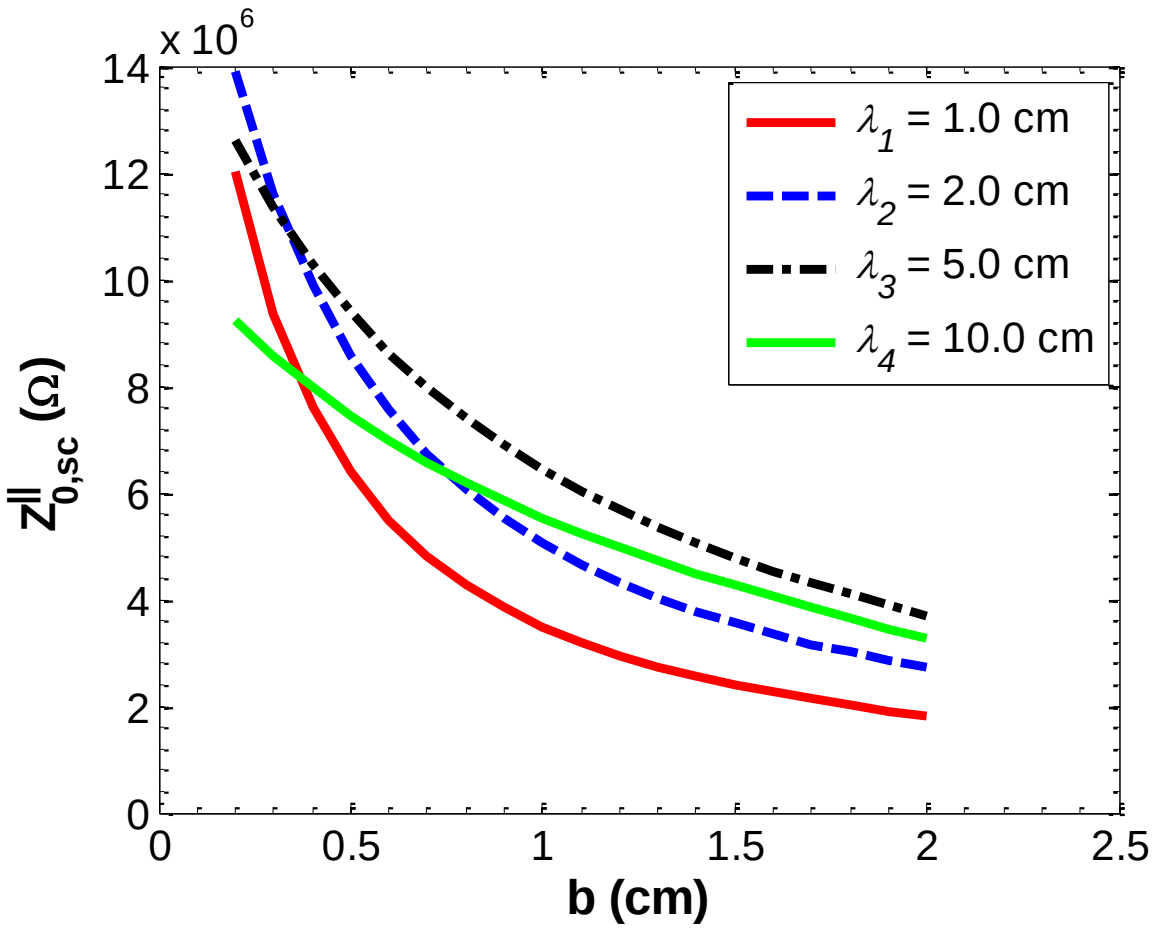

Fig. 7. Longitudinal space charge impedances of rectangular beam model with different half heights *b* inside rectangular chamber (*w* = 5.7 cm, *h* = 2.4 cm, *a* = 0.5 cm, *b* is variable).

Fig. 7 shows the calculated longitudinal space charge impedances of four perturbation wavelengths for a 20 keV $H_2^+$ beam model of rectangular cross-section inside a rectangular chamber of SIR. The parameters used in the calculations are $w$ = 5.7 cm, $h$ = 2.4 cm, $a$ = 0.5 cm, the half beam height $b$ is variable. We can see the longitudinal space charge impedances decrease with beam height *2b* for a fixed beam width *2a* due to dilutions of the volume charge densities.

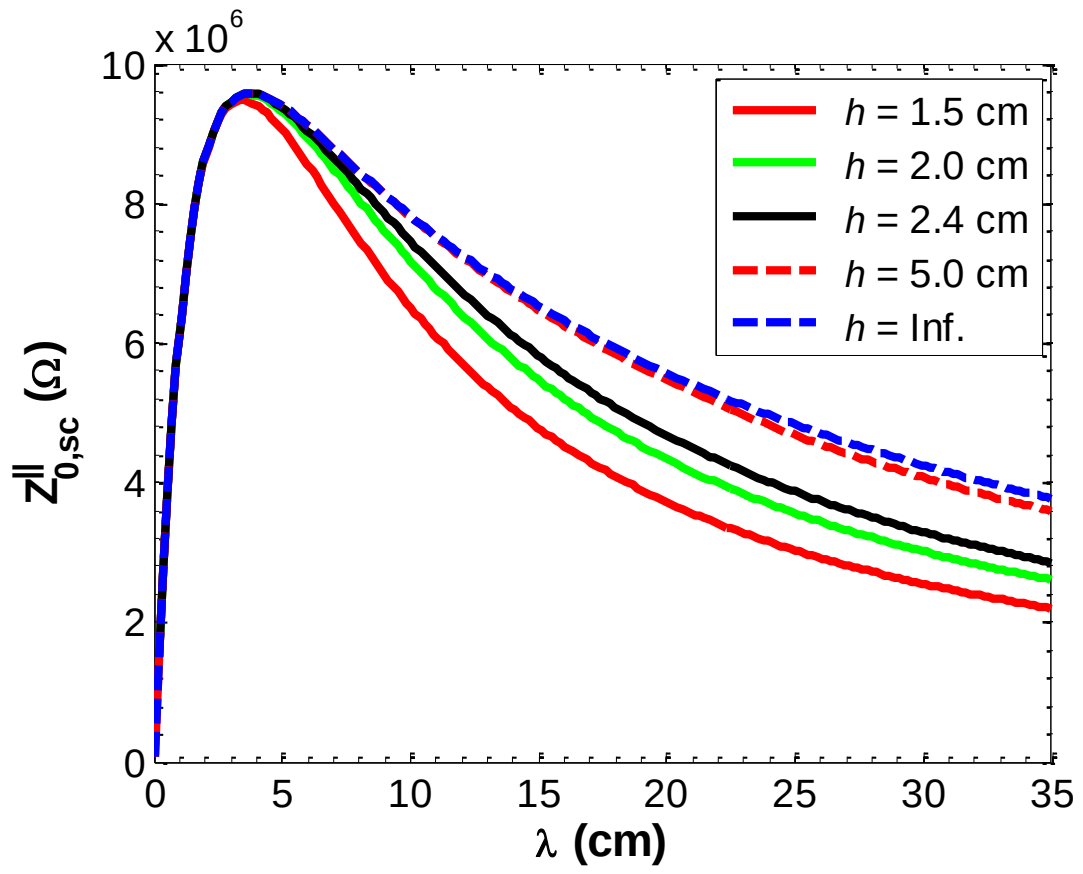

Fig. 8. Longitudinal space charge impedances of square beam model inside rectangular chamber ($w$ = 5.7 cm, $h$ is variable, $a$ = $b$ = 0.5 cm).

Fig. 8 shows the calculated longitudinal space charge impedances of a 20 keV $H_2^+$ beam model with square cross-section inside a rectangular chamber of SIR. The parameters used in the calculations are $w$ = 5.7 cm, $a$ = $b$ = 0.5 cm, the half chamber height $h$ is variable. For short wavelengths $\lambda$ < 5.0 cm, the longitudinal space charge impedances are almost independent of the changes of $h$. For longer wavelengths $\lambda$ > 5.0 cm, when $h$ > 5.0 cm, the impedances are insensitive to the changes of $h$ and are close to the limiting case of $h = \infty$ (vertical parallel plates).

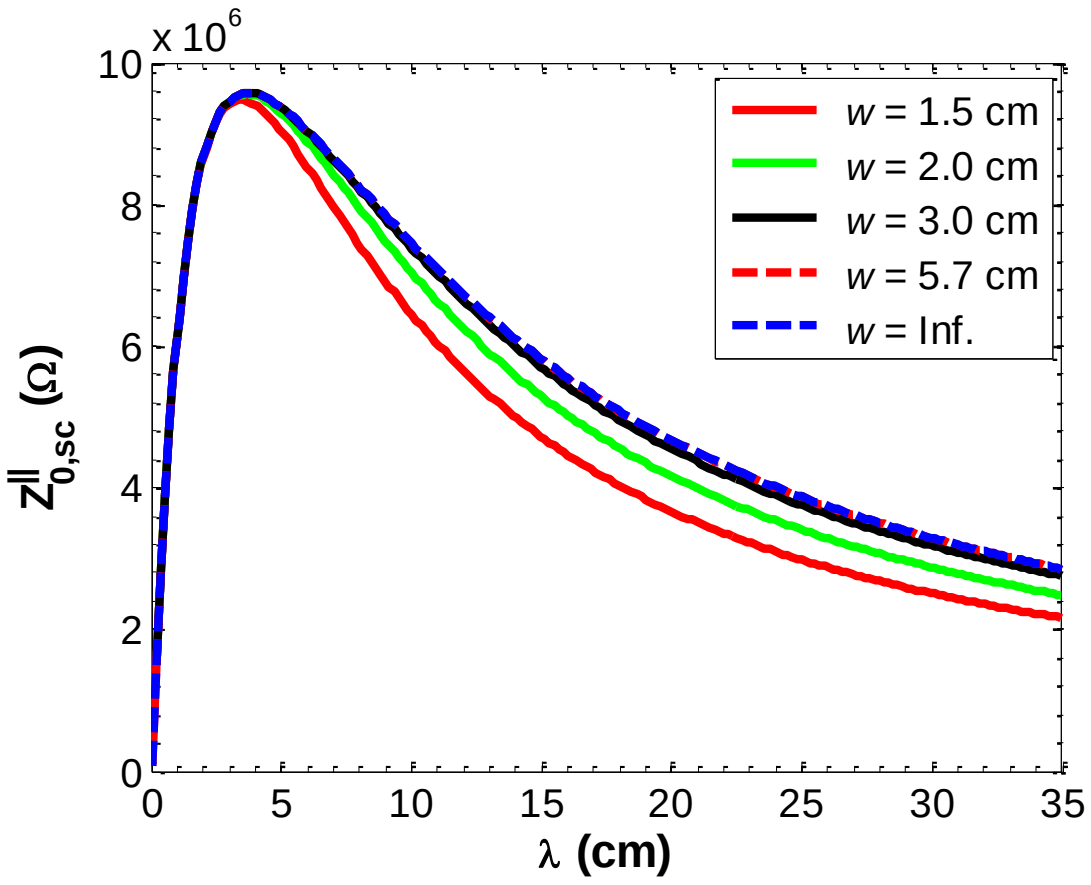

Fig. 9. Longitudinal space charge impedances of square beam model inside rectangular chamber with different half widths $w$ ($w$ is variable, $h$ = 2.4 cm, $a$ = $b$ = 0.5 cm).

Fig. 9 shows the calculated longitudinal space charge impedances of a 20 keV $H_2^+$ beam model of square cross-section inside a rectangular chamber of SIR. The parameters used in the calculations are $h$ = 2.4 cm, $a$ = $b$ = 0.5 cm, the half chamber width $w$ is variable. For short wavelengths $\lambda$ < 5.0 cm, the longitudinal space charge impedances are almost independent of the changes of $w$. For longer wavelengths $\lambda$ > 5.0 cm, when $w$ > 3.0 cm, the impedances are insensitive to the changes of $w$ and are close to the limiting case of $w = \infty$ (horizontal parallel plates).

## 6. Conclusions

We introduced a 3D space charge field model of rectangular cross-section to calculate the perturbed potentials, fields and the associated longitudinal space charge impedances. The calculated longitudinal space charge impedances are consistent with the numerical simulation results. A rectangular beam shape with $a = b = r_0$ may help to reduce the longitudinal space charge impedances compared with the conventional round beam with radius $r_0$, this result is consistent with Ref. [6] in which a planer geometry was investigated. For fixed *b(a)*, when *a(b)* increases, the longitudinal space charge impedance will decrease. The impedances of a rectangular beam inside a pair of infinity large parallel plates are also derived in this paper. Theoretical calculations demonstrate that when the transverse chamber dimensions are approximately more than 5 times of the transverse beam dimensions, the rectangular chamber of the Small Isochronous Ring (SIR) can be approximated by a pair of parallel plates. This validates the simplified boundary model of parallel plates used in the Particle-In-Cell simulation code CYCO to simulate the rectangular chamber of SIR [10].

## Acknowledgements

We would like to thank Prof. F. Marti and T. P. Wangler for their helpful guidance and discussions. This work was supported by NSF Grant # PHY 0606007.